\documentclass[journal]{IEEEtran}

\usepackage{amsmath, amssymb, amsfonts, bm}
\usepackage{graphicx}
\usepackage{float}
\usepackage{multicol}
\usepackage{mwe}
\usepackage{caption}
\usepackage[ruled,vlined]{algorithm2e}
\usepackage{subcaption}
\usepackage{hyperref}
\usepackage{xcolor}
\usepackage{dblfloatfix} 
\usepackage{acronym}

\begin{document}

\title{Low-Complexity Geometric Shaping}

\author{Ali~Mirani,
        Erik~Agrell,~\IEEEmembership{Fellow,~IEEE,}
        and~Magnus~Karlsson,~\IEEEmembership{Senior Member,~IEEE, Fellow,~OSA}
\thanks{A. Mirani and M. Karlsson are with the Department
of Microtechnology and Nanoscience, Chalmers University of Technology, SE-41296 Gothenburg,
Sweden (email: mirani@chalmers.se; magnus.karlsson@chalmers.se).}
\thanks{E. Agrell is with the Department
of Electrical Engineering, Chalmers University of Technology,  SE-41296 Gothenburg,
Sweden (email: agrell@chalmers.se).}}


\maketitle

\begin{abstract}
Approaching Shannon's capacity via geometric shaping has usually been regarded as challenging due to modulation and demodulation complexity, requiring look-up tables to store the constellation points and constellation bit labeling. To overcome these challenges, in this paper, we study lattice-based geometrically shaped modulation formats in multidimensional Euclidean space. We describe and evaluate fast and low complexity modulation and demodulation algorithms that make these modulation formats practical, even with extremely high constellation sizes with more than \mbox{$10^{28}$ points}.
The uncoded bit error rate performance of these constellations is compared with the conventional \ac{QAM} formats in the additive white Gaussian noise and nonlinear fiber channels. At a spectral efficiency of \mbox{2 bits/sym/polarization}, compared with \mbox{4-\ac{QAM}} format,  transmission reach improvement of more than 38\% is shown at the hard-decision forward error correction threshold of \mbox{$2.26\times 10^{-4}$}.
\end{abstract}

\begin{IEEEkeywords}
Optical communication, coherent receiver, multidimensional modulation format, geometric shaping, lattice.
\end{IEEEkeywords}

\IEEEPARstart{M}{uch}  effort have been put on different aspects of optical links to overcome the need for higher information throughput. Among them, the optimization of modulation formats has been an important research area \cite{winzer2012high, winzer2018fiber}.
One way to increase the transmission rate is to use higher-order modulation formats, i.e., transmitting more information bits in each channel use. However, higher-order modulation formats require more signal power to achieve a certain performance which can be a problem because fiber links are limited in transmitting power by fiber nonlinearities \cite{seimetz2009high}. Another method to improve the transmission rate in optical communication is to use constellation shaping. Shaping can be performed in two different ways, which can also be combined with each other, known as \emph{probabilistic shaping} \cite{kschischang1993optimal} and \emph{geometric shaping} \cite{forney1984efficient}. Probabilistic shaping is applied by changing the probability distribution of selecting the constellation points, whereas geometric shaping tries to optimize the position of the constellation points in the Euclidean space. Recently, probabilistic shaping has been extensively studied and shown to provide rate adaptability and energy efficiency gains \cite{cho2019probabilistic,fehenberger2016probabilistic}. However, its need for a distribution matcher increases the system complexity. Also geometric shaping has been studied, although less, but used to design nonlinear tolerant and multidimensional constellations \cite{chen2019polarization,jones2018deep,el2017multidimensional}.

Exploiting multidimensional modulation formats brings about two independent performance gains known as \mbox{\emph{coding gain}} and \emph{shaping gain}. Coding gain comes from the fact that constellation points can be packed more densely in multidimensional space, and the shaping gain is achieved when the constellation boundary is closer to a hypersphere. The maximum shaping gain that can be obtained by a spherical boundary in high dimensional space is 1.53 dB compared to a cubic boundary \cite{forney1989multidimensional}.

The electromagnetic field has 4 degrees of freedom \cite{karlsson2010four} and combined with coherent detection schemes, receivers with improved sensitivity can be utilized to compensate for channel distortions, i.e., polarization drift and nonlinear effects \cite{ip2008coherent}.
Furthermore, taking advantage of wavelengths, time slots, and spatial modes/cores in a fiber link can increase the available degrees of freedom and increase data transmission rates. Typically, these dimensions have been modulated independently in fiber optic communications, but there are studies that show that optimizing formats in many dimensions will improve the performance of the system. 

Modern research on multidimensional modulation formats in optical communication systems dates back to 2009 when the possibility of optimizing the constellations in the \mbox{4-dimensional} space formed by the optical field was demonstrated \cite{agrell2009power}. High-dimensional modulation formats were then optimized by sphere packing to find the most power-efficient and spectral-efficient modulation formats \cite{karlsson2011power}. Later on, there were even more research on employing higher dimensions such as 8 and 24.
In \cite{koike2013eight}, 8-dimensional lattice modulation formats were simulated based on spherical cut \cite{koike2009sphere} with 128 and 256 constellation points. 
Also, by combining pulse-position modulation with polarization-switched \ac{QPSK}, an 8-dimensional modulation format was experimentally investigated in \cite{eriksson2014biorthogonal} showing significant improvement over polarization multiplexed \ac{QPSK}.
In \cite{lu2017optimized} and \cite{lu201724}, 8- and 24-dimensional modulation formats using temporally adjacent time slots were used to transmit data over short-reach intensity modulated links.
In \cite{millar2014high}, multidimensional modulation formats in coherent optical systems were demonstrated by spherical-cut constellations and block-coded modulation. The advantages of combining multidimensional shaping with non-binary coded modulation were shown in \cite{matsumine2019short}.
Optimized 8-dimensional modulation format based on polarizations and time slots was compared with polarization multiplexed \ac{BPSK} to show its improved nonlinear performance in \cite{shiner2014demonstration}.
In \cite{forney1989multidimensional2}, multidimensional constellations were investigated based on lattices called \mbox{\textit{\ac{VC}}} and were suggested as coded modulation schemes with higher shaping gains than generalized cross constellations indicated in \cite{forney1989multidimensional}. 
Also, the densest lattices for the sphere packing problem in different dimensions with fast and low-complexity modulation and demodulation algorithms were studied in \cite{conway1983fast} and \cite{conway1982fast}.

In this paper, we evaluate the performance of coherent fiber links with ultra-large Voronoi constellations with more than $10^{28}$ points, which has never been done before to our knowledge. The low-complexity algorithms are numerically compared with \ac{ML} detection and shown to perform almost equivalently at high \ac{SE} or high \ac{SNR}. Low-complexity natural binary and quasi-Gray constellation bit labelings are used to map bits to constellation points and their performances are compared. The asymptotic power efficiency of lattices for different dimensions and \ac{SE}s, as well as the optimum choice of Voronoi boundary are studied. The uncoded \ac{BER} performance of these \ac{VC}s are then examined in the \ac{AWGN} and nonlinear fiber-optic channels.

The remainder of the paper is organized as follows. In section II, we describe the lattice theory, lattice properties, and how to use them for finite constellations.
Fast modulation and demodulation algorithms for lattice-based \ac{VC}s are then presented in section II.
In section III, figures of merit for comparing modulation formats at different dimensions are defined and numerical simulations are applied to demonstrate these concepts. We finally give the performance of the \ac{VC}s in section IV in both the \ac{AWGN} and the nonlinear fiber-optic channels. The conclusion of this paper is presented in section V.

\section{Preliminaries}
\label{prelim}
In this section, we introduce the basics of lattice theory, lattice properties, constellations and their construction based on lattices.

In general, lattices are periodic structures of points in \mbox{$N$-dimensional} Euclidean space. Mathematically, lattices can be described and constructed using a set of linearly independent column vectors
$\bold g_1,\bold g_2,\cdots, \bold g_l \in \mathbb{R}^N$ 
in $N$-dimensional space where $\mathbb{R}$ is the set of real numbers and $l\leq N$. Linear combination of these vectors with integer coefficients creates the lattice points which are then defined by
\begin{align}
    \label{lattice}
    \Lambda = \left \{ z_1\bold g_1+\cdots+z_l\bold g_l \,\Big|\, z_i\in\mathbb{Z},i\in \left \{1,\cdots, l\right \} \right \}.
\end{align}
In this paper, we only consider full-rank lattices, i.e., \mbox{$l=N$}. In order to describe lattices in a more compact way, the square matrix $\bold G = \{\bold g_1,\,\bold g_2,\,\cdots,\,\bold g_N\}\in \mathbb{R}^{N\times N}$ is defined as the \textit{generator matrix} for the lattice $\Lambda$ which has the property of $\text{det}(\bold G)\neq 0$. Therefore, the lattice points description in (\ref{lattice}) can be rewritten as
$\Lambda = \left \{\bold G \bold z\, |\, \bold z \in \mathbb{Z}^N\right \}$. It should be noted that the generator matrix is not unique and different generator matrices can create the same lattice points.

In communication systems, information bits are mapped to symbols that are sent every $T_s$ seconds and modulate a carrier which are then transmitted through a channel. Symbols are selected from a set of $M$ elements, the \emph{constellation}, $\mathcal{C}=\{\bold c_0,\bold c_1, ...,\bold c_{M-1}\}$. Constellations can be described in an $N$-dimensional space and therefore each symbol ($\bold c_i$) is represented by an $N$-dimensional vector. In this work, an equal probability of generating each symbol from the constellation is considered. Therefore, the entropy of this source or average information of each symbol is $m = \text{log}_2 M$ \textit{bits}, and the information rate of the source is equal to $R_b = \frac{m}{T_s}$ \textit{bits per second}.

The average energy per symbol is defined as
\begin{align}
    E_s = \frac{1}{M}\sum_{i=0}^{M-1} \|\bold c_i \|^2=m\cdot E_b,
\end{align}
where $E_b$ is the average energy per bit. The average power is calculated as $P = {E_s}/{T_s}$.

In order to use the (infinite) set of points in a lattice in communications, we need to select a finite set of points as a constellation. In other words, we need to cut a constellation from the lattice. There are different ways of doing this, e.g., spherical or cubic cuts \cite{millar2014high, karlsson2012spectrally}, but what is interesting for us in this paper is the \emph{Voronoi cut}.
\begin{figure}[t]
    \begin{center}
        \includegraphics[width=0.4\textwidth]{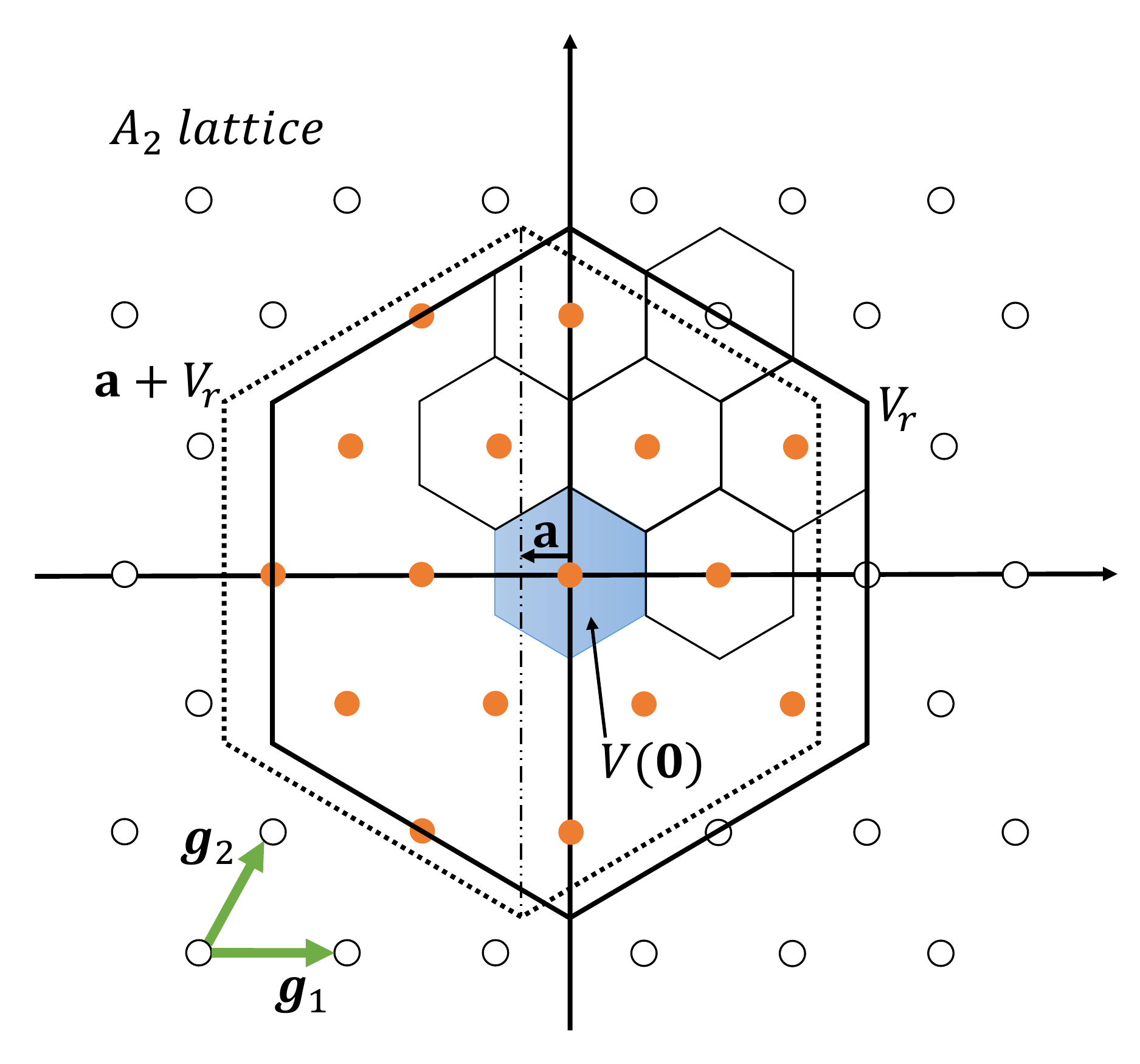}
        \caption{Properties for $A_2$ lattice. The vectors $\bold g_1=[1\,\, 0]^T$ and \mbox{$\bold g_2=[1/2\,\, \sqrt{3}/2]^T$} show the basis vectors of the lattice. Circular points show the lattice points, small hexagons are the Voronoi regions around the lattice points and $V(\bold 0)$ is the Voronoi region of the lattice. The filled points are the selected constellation points with a Voronoi cut.  The boundary $V_r$ indicates the scaled Voronoi region with $r=4$, containing $r
       ^2=16$ points. The shift vector $\bold a$ is chosen to remove the lattice points on the border of $V_r$. }
        \label{voronoi}
    \end{center}
\end{figure}%
The Voronoi region around a lattice point $\bold x$, known as \mbox{$V(\bold x)$}, is defined as the set of points in $\mathbb{R}^N$ which are closer to point $\bold x$ than any other points of the lattice, i.e.,
\begin{align}
    V(\bold x) = \left \{\bold y \in \mathbb{R}^N\, \Big|\, \|\bold y-\bold x\|\leq \|\bold y-\bold w\|\,,\forall \bold w \in \Lambda\right \}.
\end{align}
It should be noted that the all-zero vector, $\bold 0$, is always a lattice point and the Voronoi region $V(\bold 0)$ is called the Voronoi region of the lattice.

To select $M=r^N$ lattice points from the lattice $\Lambda$ with a Voronoi cut, where $r=1,2,3,\cdots$, we consider the scaled Voronoi region of the lattice $r\Lambda = \left \{ r\bold x\,|\,x\in \Lambda\right\}$ and call it $V_r$. This Voronoi region is  an $r$ times scaled version of the $V(\bold 0)$ in lattice $\Lambda$. The $M$ constellation points are defined as the set of \mbox{$\mathcal{C}_{\Lambda}(r,\bold a)=\left\{\bold x-\bold a\,\Big|\,\bold x\in \Lambda \cap (\bold a + V_r) \right\}$},
where \mbox{$\bold a\in \mathbb{R}^N$} is a \emph{shift vector} chosen to avoid any lattice points on the boundary of $\bold a +V_r$. The choice of $\bold a$ will affect the average energy of the constellation, and the best choice of it will be discussed in a later section.

The concepts of basis vectors, infinite lattice, Voronoi regions, scaled and shifted Voronoi regions, and finite constellation points are all exemplified in Fig. \ref{voronoi} for the two-dimensional hexagonal lattice $A_2$.

The hexagonal lattice is the densest lattice in two-dimensional space. Finding the densest structure in a given dimension is known as a sphere packing problem, and is notoriously difficult. Lattices that are used in this paper are the cubic lattice ($\mathbb{Z}^N$), the hexagonal lattice ($A_N$, known to be densest in $N=2$ and $N=3$), the checkerboard lattice ($D_N$, densest in $N=4$), the
Gosset lattice ($E_8$, densest in $N=8$), and the Leech lattice ($\Lambda_{24}$, densest in $N=24$). For a more in-depth discussion of these lattices and their properties, see \cite{conway2013sphere}.

\section{Voronoi constellation design}
In this section, we design \ac{VC}s of various dimensions $N$ and sizes $M$. This is not done by tabulating all constellation points, as in traditional geometric shaping, but by algebraic descriptions of the constellations. Specifically, a Voronoi constellation $\mathcal{C}_\Lambda(r,\bold a)$ is fully determined by the generator matrix $\boldsymbol{G}$, the scale factor $r$, and the shift vector $\bold a$.

\subsection{Modulation}
After introducing the infinite lattices and creating finite constellations using Voronoi cuts, the question that remains is how to select a constellation point with low computational complexity to transmit the information bits \cite{conway1983fast}.

Suppose that the constellation based on lattice $\Lambda$ is \mbox{$\mathcal{C}_\Lambda (r,\bold a)=\{\bold c_0,\bold c_1,\cdots,\bold c_{M-1}\}$} with $M=r^N$ constellation points. Using $m = \log_2 M$ bits, one of the constellation points can be uniquely selected. To avoid storage problems, we want to avoid storing constellation points in a look-up table, but generate each required constellation point based on its index each time it is needed. Therefore, $m$ bits are mapped to a constellation index $K\in\{0,1,\cdots,M-1\}$. Then, following Algorithm 1 \cite{conway1983fast}, a \ac{VC} point for the index $K$ can be selected.

\begin{algorithm}[b]
\SetAlgoLined
\KwResult{$K\rightarrow \bold c$}
 1: $ K\xrightarrow{\text{changing base}} ( k_0 k_1\cdots k_{N-1} )_r,\quad 0\leq k_i\leq r-1$\\
 2: $\bold x=\sum_{i=1}^N k_i \bold g_i$\\
 3: $\bold w = (\bold x-\bold a)/r$\\
 4: $\bm{\lambda} = \text{CPA}(\bold w),\quad \bm{\lambda}\in\Lambda$\\
 5: $\bold c = \bold x - r\bm{\lambda}-\bold a,\quad\bold c\in\mathcal{C}_\Lambda(r,\bold a)$
 \caption{\mbox{Modulation (index to coordinates)}}
\end{algorithm}
In the first line of the algorithm, $K$ is represented in base $r$ using \mbox{$K=k_0r^{N-1}+k_1 r^{N-2}+\cdots+k_{N-1} r^0$}. Then, in the subsequent lines, a point in the lattice is selected and finally mapped to the region where the constellation is considered based on $\mathcal{C}_\Lambda (r,\bold a)$. In the fourth line, the \ac{CPA} is used to find the closest lattice point to $\bold w$. The \ac{CPA} is described in \cite{conway1982fast} and we will discuss it briefly in section \ref{CPA-sec}.

\subsection{Demodulation}
After selecting a constellation point ($\bold c$) and transmitting it over the channel, in the receiver side, the received samples \mbox{($\bold y\in\mathbb{R}^N$)} need to be mapped back to a probable \ac{VC} point to find its index and the transmitted bits. This procedure is described in Algorithm 2 \cite{conway1983fast}.
\begin{algorithm}[t]
\SetAlgoLined
\KwResult{$\bold y\rightarrow K$}
 1: $\bold w = \bold y + \bold a$\\
 2: $\bm{\lambda} = \text{CPA}(\bold w),\quad \bm{\lambda}\in\Lambda$\\
 3: $\bold k' = \bold G^{-1}\bm{\lambda},\quad \bold k'=[k'_0,k'_1,\cdots,k'_{N-1}]^T$\\
 4: $\bold k = \bold k'\pmod{r},\quad \bold k=[k_0,k_1,\cdots,k_{N-1}]^T$\\
 5: $( k_0 k_1\cdots k_{N-1} )_r\xrightarrow{\text{changing base}}K$
 \caption{\mbox{Demodulation (received samples to index)}}
\end{algorithm}

\begin{figure}[h]
\begin{subfigure}{.5\textwidth}
  \centering
  \includegraphics[width=.8\linewidth]{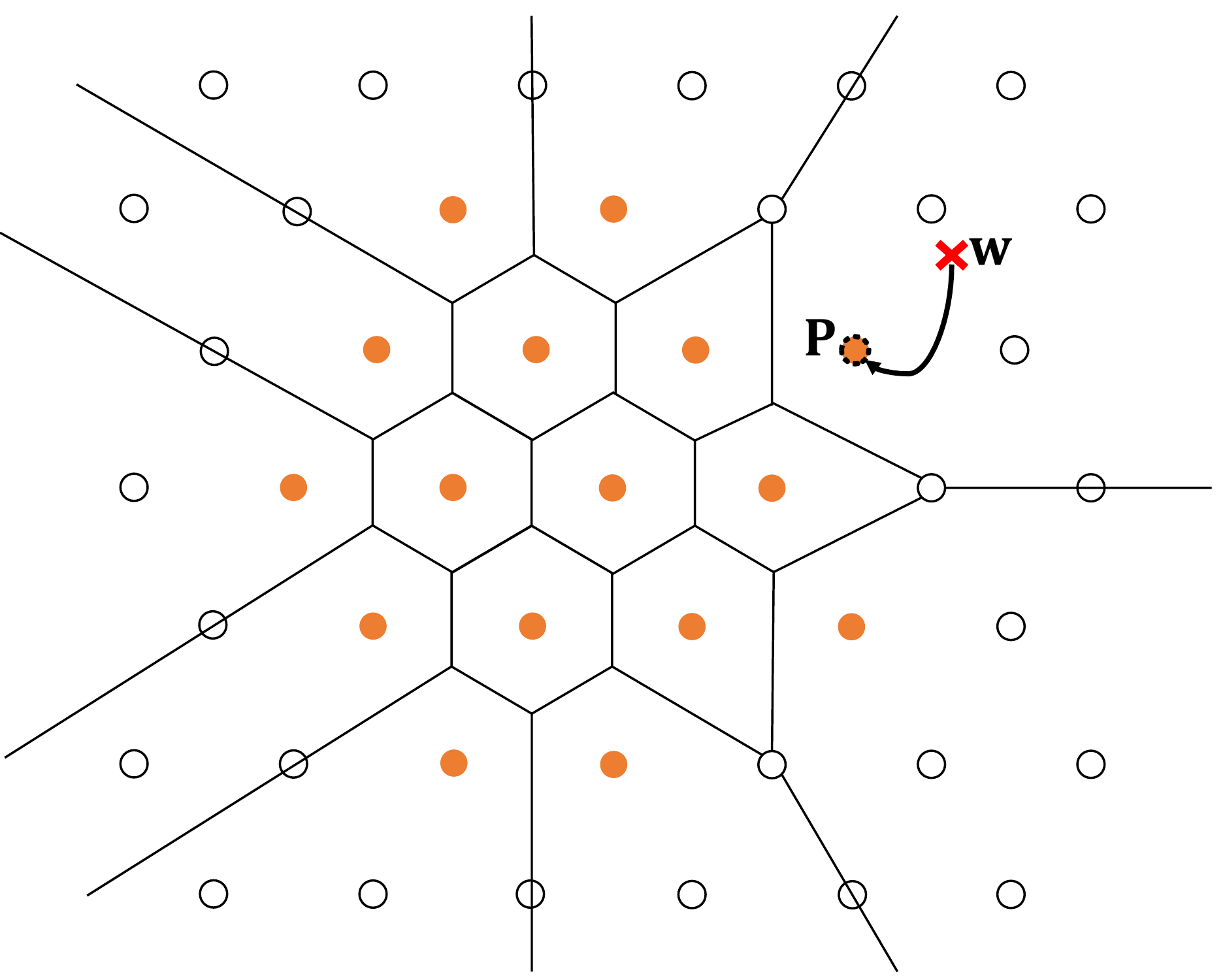}
  \caption{Maximum likelihood (ML) detection.}
  \label{ML}
\end{subfigure}
\begin{subfigure}{.5\textwidth}
  \centering
  \includegraphics[width=.8\linewidth]{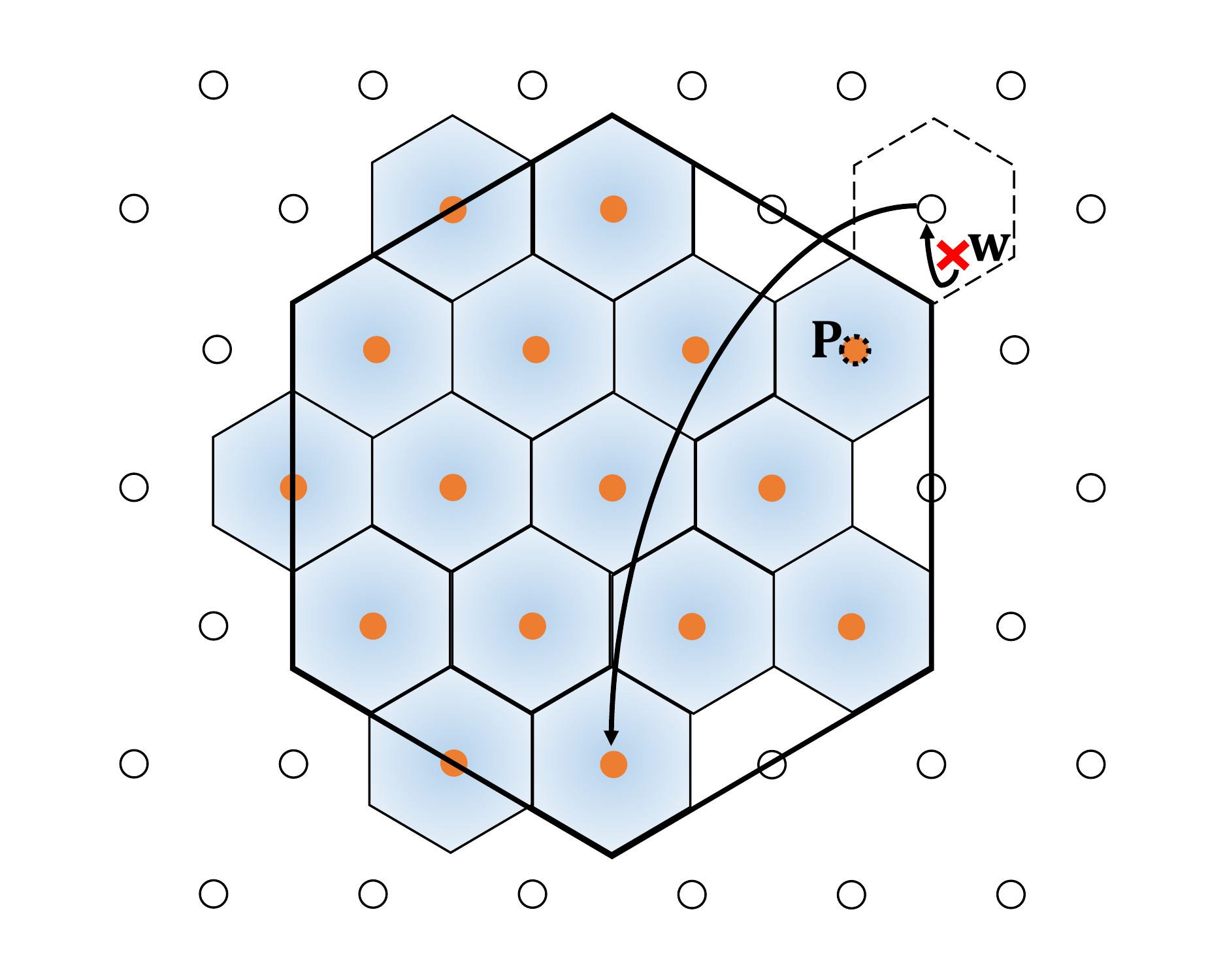}
  \caption{Detection using Algorithm 2.}
  \label{LD}
\end{subfigure}
\caption{Detection methods in the receiver.}
\label{ML_LD}
\end{figure}

First, the received point is shifted from the finite constellation coordinates to the infinite lattice coordinates. Then, the closest lattice point is found by \ac{CPA} and the index for this lattice point is calculated. Since this point can be outside the constellation area, it is moved back to a point inside the constellation region by the modulo operation (this procedure is shown in Fig. \ref{LD} by arrows). Finally, the base of the index is changed and bits can be recovered in the end.

The demodulation algorithm is almost the reverse procedure of the modulation algorithm and their core part is the \ac{CPA}. Therefore, the complexity of the modulation and the demodulation algorithms are similar. This is one of the advantages of using the lattice structures as the constellation in transmission systems.

\subsection{Closest point algorithm}
\label{CPA-sec}

The \ac{CPA} which is used in the modulation and demodulation procedure is the main part of these algorithms and it provides an optimum way to find the closest point of an infinite lattice to a point in $\mathbb{R}^N$. The \ac{CPA} was described in \cite{conway1982fast} for different lattices such as $\mathbb{Z}^N, A_N, D_N,\text{ and } E_8$. To implement \ac{CPA} for $\Lambda_{24}$, we use the information in \cite{conway1984voronoi} and \cite{adoul1988nearest} and the construction based on extended Golay code and $D_{24}$ lattice. Usually, a more complicated lattice can be constructed using a union of cosets (shifted lattices) of other simpler lattices. Therefore, finding the closest lattice point in these complicated lattices will be simplified to applying \ac{CPA} to the simpler cosets and finally comparing their results to find the closest point. For instance, $A_2$ can be viewed as two rectangular lattices (scaled $\mathbb{Z}^2$ lattices) and $E_8$ is the union of two cosets of $D_8$. Consequently, the \ac{CPA} is applied over these cosets and their results are compared to find the closest point.
The \ac{CPA} also finds the closest lattice point in an infinite lattice to a point in $\mathbb{R}^N$ with few comparisons which depends on the number of dimensions regardless of the infinite number of lattice points.
The \ac{CPA} independence from the number of lattice points is one of the most important advantages of using lattice-based geometrically shaped constellations compared to other shaping methods.

\subsection{The constellation shift vector}
For every choice of constellation shift vector $\bold a$, there exists in general a different \ac{VC} for the same choices of the lattice and \ac{SE}.
In order to find the best shift vector to minimize the constellation energy, an iterative algorithm was suggested in \cite{conway1983fast} which can converge in a few iterations. However, if the number of constellation points is high such as in constellation taken from $\Lambda_{24}$, this iterative algorithm will be complex.

\begin{figure}[h]
    \begin{center}
        \includegraphics[width=0.5\textwidth]{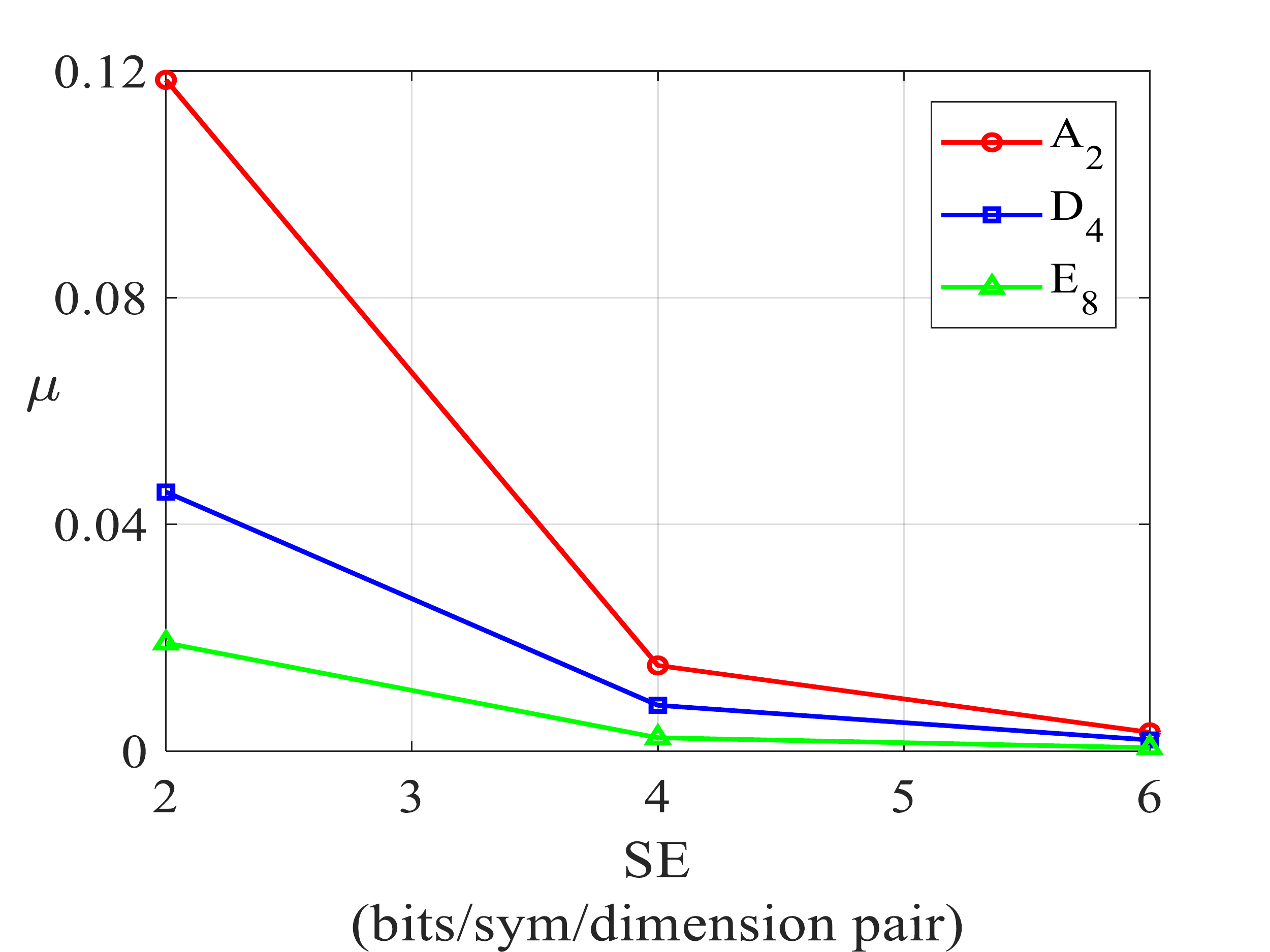}
        \caption{Effect of random selection of the shift vector on the constellation energy.}
        \label{a}
    \end{center}
\end{figure}

In Fig. \ref{a}, we show that when the number of constellation points increases, we can ignore the optimization of finding the best shift vector, i.e., for large constellations, choosing a shift vector randomly will give almost the same constellation average energy ($E_s$) as optimizing it. To show this concept, we define
\begin{align}
    \mu = \frac{\mathbb{E}_{\bold a}\left[E_s\right]-E_{s,opt}}{E_{s,opt}}
\end{align}
where $\mathbb{E}_{\bold a}$ stands for expectation with respect to choosing $\bold a$ uniformly in $V(\bold 0)$. Also, $E_{s,opt}$ is the energy of the constellation generated by the iterative algorithm in \cite{conway1983fast}.
Figure \ref{a} shows that the normalized difference of energy, between randomly selected and optimized choice of the shift vector, decreases as the number of constellation points increases.

\section{Figures of merit}
\label{figOfMerit}
In this section, parameters to study and compare different multidimensional modulation formats are introduced.

\subsubsection{Spectral efficiency}
\label{SE-txt}
For a constellation with $M$ separate symbols in $N$ dimensions, the \ac{SE} is calculated as 
\begin{align}
    \label{eq:SE}
    SE = \frac{\log_2 M}{N/2}=2\log_2 r
\end{align}
and has the unit of $\textit{bits per symbol per dimension pair}$. In fiber-optic communication, usually, the polarization of light is considered as a pair of dimensions because each polarization can support two dimensions of in-phase and quadrature components. If Nyquist pulse shaping has been used in the frequency domain to transmit symbols, the unit of \ac{SE} can also be considered as $\textit{bits per second per Hertz}$ which shows how efficient the frequency spectrum is being used \cite{karlsson2016multidimensional}. 

Based on Eq. (\ref{eq:SE}), the \ac{SE} is related to the scaling factor ($r$) of the Voronoi region. This shows that any \ac{SE}s can be achieved by changing $r$ for every lattice $\Lambda$. In this work, to simplify bit mapping, we focus on scaling factors that are powers of 2, i.e., $r = 2, 4, 8, 16$.

\subsubsection{Sensitivity penalty}
\label{PE}
\begin{figure}[t]
    \begin{center}
        \includegraphics[width=0.5\textwidth]{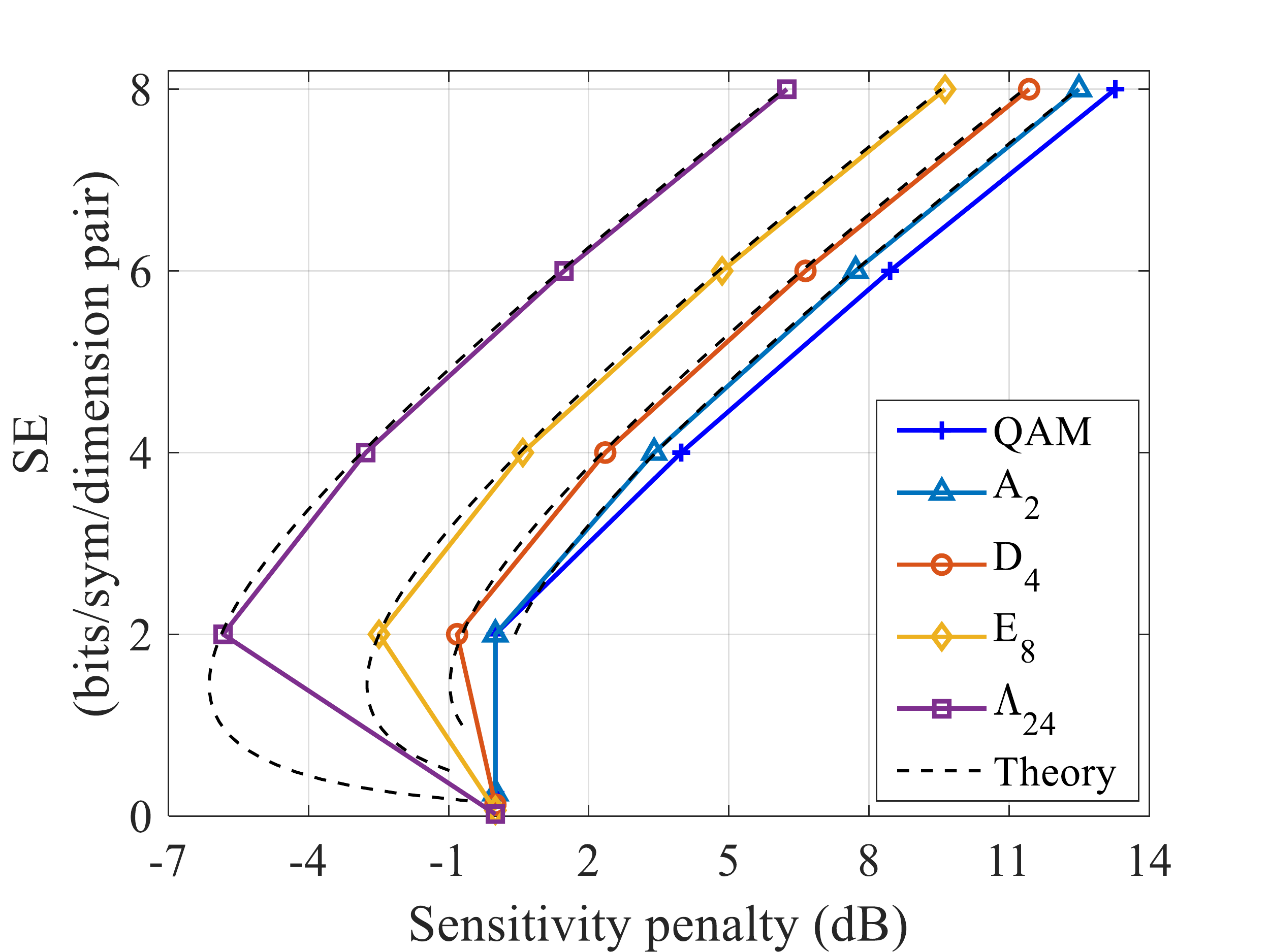}
        \caption{The \ac{SE} vs. sensitivity penalty for lattice base multidimensional modulation formats. The dashed lines are calculated using Eq. (23) in \cite{agrell2009power}.}
        \label{SEvsSP}
    \end{center}
\end{figure}

For every modulation format, $d_{min}$ is the minimum Euclidean distance between two symbols of the constellation. For high \ac{SNR} and the \ac{AWGN} channel, the probability of symbol error can be approximated using the union bound and its dominant terms by
$
 \text{erfc}\left({d_{min}}/\left({2\sqrt{N_0}}\right)\right)
$
where \mbox{$\text{erfc}(x)=\left({2}/{\sqrt{\pi}}\right)\int_x^\infty \exp{\left({-t^2}\right)}dt$} is the complementary error function and $N_0/2$ is the variance of the Gaussian noise in each dimension. Therefore, the \ac{SER} is a monotonically decreasing function of
\begin{align}
    \frac{d_{min}^2}{4N_0}=\frac{P}{R_b N_0}\gamma=\frac{E_b}{N_0}\gamma
\end{align}
where $\gamma = d_{min}^2/(4E_b)$ is defined as the asymptotic power efficiency because for a given \ac{SER}, the required power is proportional to $1/\gamma$. The parameter $\gamma$ describes the constellation geometry and is often given in dB. If different modulation formats are compared at the same power and bit rate, $\gamma$ shows the power gain for high \ac{SNR} with respect to \ac{BPSK}, \ac{QPSK} and dual polarization \ac{QPSK} modulation formats because for these constellations $\gamma = 0\ \rm{dB}$. The sensitivity penalty is also defined as $1/\gamma$ and it shows the performance penalty compared to \ac{BPSK}, \ac{QPSK} and dual polarization \ac{QPSK} for high \ac{SNR} \cite{agrell2009power}. The  sensitivity penalty comparison of different \ac{VC}s with respect to \ac{QAM} are shown in Fig. \ref{SEvsSP}.

For multidimensional modulation formats, as the \ac{SE} increases, the asymptotic power efficiency ($\gamma$) becomes proportional to the product of two important gains over the cubic lattice called the shaping gain ($\gamma _s$) and coding gain ($\gamma _c$) \cite{forney1989multidimensional}. Shaping gain shows how spherical the boundary of the constellation is, and in this work, we use a Voronoi cut. The coding gain indicates the density of the lattice points. These gains for different lattices are shown in Table \ref{shape_code_gain} and their mathematical definitions were presented in \cite{forney1989multidimensional}.

\subsubsection{Symbol and bit error rate}
For equiprobable constellations with \ac{ML} estimation, using pairwise error probability and union bound, the \ac{SER} is bounded by 
\begin{align}
    \text{SER} \leq \frac{1}{M}\sum_{i=0}^{M-1}\sum_{j\neq i}\frac{1}{2}\text{erfc}\left(\frac{d_{ij}}{2\sqrt{N_0}}\right)
\end{align}
where $d_{ij}$ is the Euclidean distance between $\bold c_i$ and $\bold c_j$.
For high \ac{SNR}, the terms with $d_{ij}=d_{min}$ dominate, and for lattices, all constellation points have at least one neighbor at distance $d_{min}$ \cite{benedetto1999principles}. Therefore, the \ac{SER} is bounded by
\begin{align}
    \label{SER_approx}
    \frac{1}{2}\text{erfc}\left(\frac{d_{min}}{2\sqrt{N_0}}\right)\leq \text{SER}\,\widetilde{\leq}\, \frac{\bar{\tau}}{2}\text{erfc}\left(\frac{d_{min}}{2\sqrt{N_0}}\right)
\end{align}
which $\widetilde{\leq}$ means an approximate upper bound that approaches the true value as $N_0$ goes to zero. Also, \mbox{$\bar{\tau}=({1}/{M})\sum_{i=0}^{M-1} \tau_i$} which $\tau_i$ is the number of neighbours at the  minimum distance from symbol $\bold c_i$. Equation (\ref{SER_approx}) shows that two important factors for \ac{SER} of lattices are $d_{min}$ and average of number of closest neighbors which for high-dimensional lattices can be very high \cite{benedetto1999principles}. 

The \ac{BER} depends on the bit labeling of the symbols as well. For dense packing high dimensional lattices, there is no Gray mapping solution to optimize the \ac{BER} performance. In this paper, instead of looking for the best mapping, we looked at two fast and low-complexity methods which we call natural binary and quasi-Gray constellation bit labeling. Natural binary labeling is the direct transformation of the symbol index ($K$) to its base-2 representation. Quasi-Gray labeling means transforming the symbol indexes in each dimension ($k_i$) to a binary Gray code.
\begin{table}[t] 
    \centering
    \caption{Shaping and coding gain of multidimensional lattices \cite{forney1989multidimensional2,benedetto1999principles}}
    \begin{tabular}[t]{lccccccccccc}
        \hline
        Lattice&\,&\,&$\mathbb{Z}^N$&\,&$A_2$&\,&$D_4$&\,&$E_8$&\,&$\Lambda_{24}$\\
        \hline\hline
        $\gamma _c$ (dB) &\,&\,&0&\,&0.62&\,&1.51&\,&3.01&\,&6.02\\
        $\gamma _s$ (dB) &\,&\,&0&\,&0.17&\,&0.37&\,&0.65&\,&1.03\\
        \hline
    \end{tabular}
    \label{shape_code_gain}
\end{table}%

\section{Results}
In this section, we present the simulation results applying the \ac{VC}s in the \ac{AWGN} and nonlinear fiber channel. The results are shown as the performance of these modulations in terms of uncoded \ac{BER} and \ac{SER}.

\subsection{Additive white Gaussian noise channel}
In this part, the bit labeling methods, detection algorithms in the receiver, and the performance of the \ac{VC}s with \mbox{Algorithms 1 and 2} are compared over the \ac{AWGN} channel with respect to the equivalent \ac{QAM} formats with \ac{ML} detection.

\begin{figure}[h]
    \begin{center}
        \includegraphics[width=0.5\textwidth]{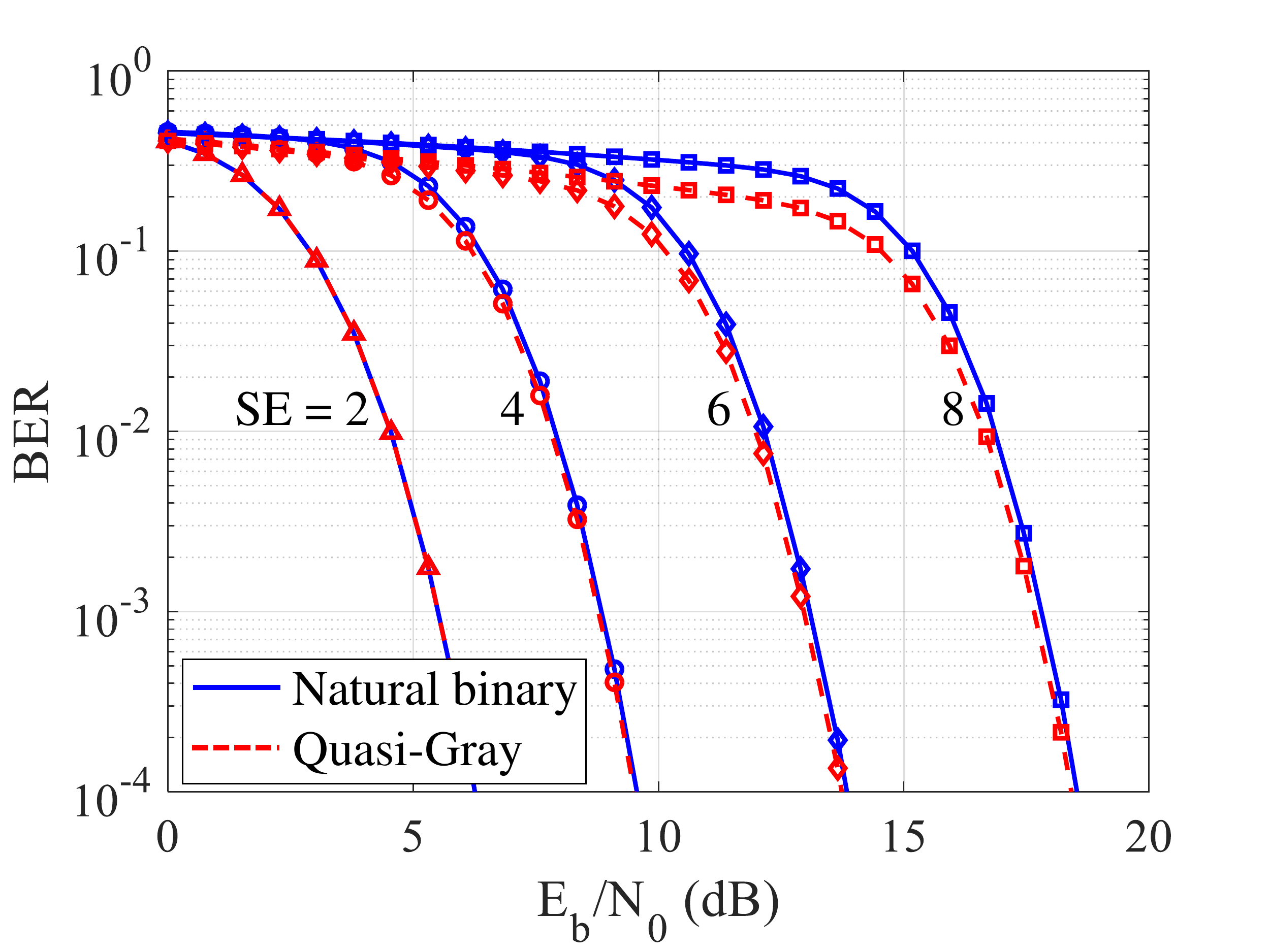}
        \caption{Natural binary vs. quasi-Gray bit labeling for Leech lattice at different \ac{SE}s over \ac{AWGN} channel.}
        \label{Gray}
    \end{center}
\end{figure}

To compare the natural binary and quasi-Gray constellation bit labeling, we simulate their \ac{BER} performance over the \ac{AWGN} channel and use them to encode and decode the transmitted and received \ac{VC} symbols.
The simulation setup includes a transmitter which generates multidimensional symbols and a multidimensional \ac{AWGN} channel \mbox{$\mathcal{N}(0,\frac{N_0}{2}\mathbf{I}_{N\times N})$}. In the receiver, we use \mbox{Algorithm 2} to estimate the \ac{VC} symbols.
At a \mbox{$SE$ = 2 bits/sym/dimension pair}, these two methods create the same symbol labeling for the constellation points. However, for higher \ac{SE}s, quasi-Gray performs better as shown in Fig. \ref{Gray} for the Leech lattice ($\Lambda_{24}$).

\begin{figure}[h]
    \begin{center}
        \includegraphics[width=0.5\textwidth]{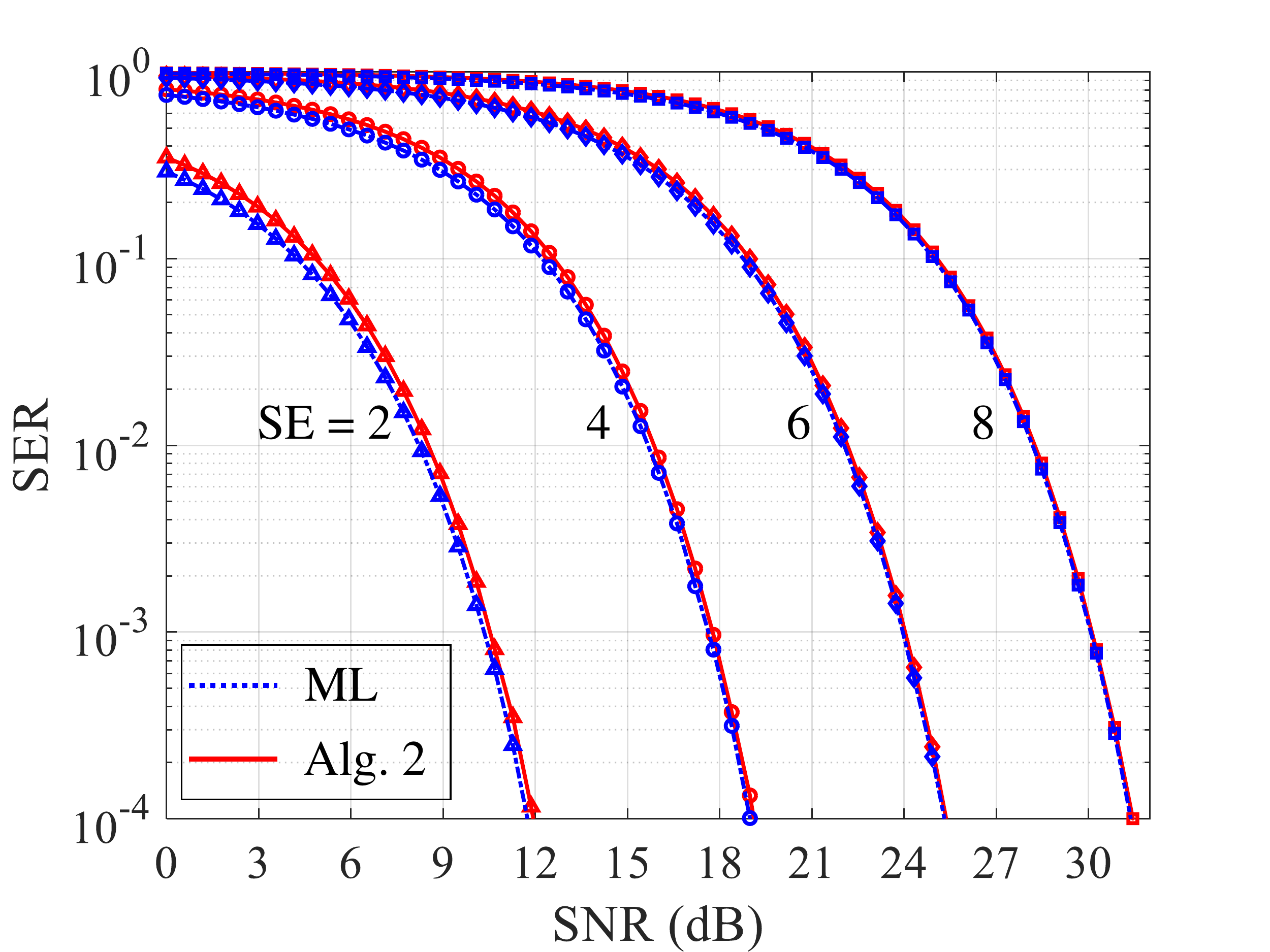}
        \caption{The \ac{ML} estimation vs. \mbox{Algorithm 2} performance for \ac{QAM} formats at different \ac{SE}s over \ac{AWGN} channel.}
        \label{SER_ML_LD}
    \end{center}
\end{figure}

In the receiver, an alternative detection scheme for \ac{VC}s is the \ac{ML} estimation which is the optimum detection rule for constellations with uniform probability distribution of symbols.
The \ac{ML} estimation compares the received information with all possible constellation points and finds the one which has the minimum Euclidean distance from the received point. The \ac{ML} estimation requires a lookup table the size of the constellation cardinality. Therefore, for our \ac{VC}s, \ac{ML} detection can be extremely time consuming for the constellation sizes shown in Table \ref{size}. However, the fast and low-complexity performance of \mbox{Algorithm 2} comes at the expense of suboptimal behaviour, which is illustrated in \mbox{Fig. \ref{ML_LD} and \ref{SER_ML_LD}}.
\begin{table}[h]
\centering
\caption{Constellation size ($M$) in different dimensions and \ac{SE}s}
\begin{tabular}[t]{lcccc}
\hline
SE&2&4&6&8\\
\hline\hline
$A_2$ &4&16&64&256\\
$D_4$ &16&256&4,096&65,536\\
$E_8$ &256&65,536&16,777,216&4,294,967,296\\
$\Lambda_{24}$ &16,777,216&$\approx 2.8\times 10^{14}$&$\approx 4.7\times 10^{21}$&$\approx 7.9\times 10^{28}$\\
\hline
\end{tabular}
\label{size}
\end{table}%

In Fig. \ref{ML}, the decision regions for the \ac{ML} estimation are shown. If point $\bold P$ is transmitted and $\bold w$ is received, since it is still in the decision region of point $\bold P$, it will be detected correctly. However, in Fig. \ref{LD}, using \mbox{Algorithm 2}, the received point will be mapped to an incorrect constellation point. This situation only happens when the received vector $\bold w$ is outside the Voronoi region of the transmitted vector ($V(\bold P)$) and also outside the scaled Voronoi region ($V_r$). Therefore, when the number of constellation points increases, the ratio between the boundary constellation points and the total number of points decreases, and these errors will be less likely. Also, in high \ac{SNR} regimes, the received points are more probable to remain in the Voronoi region of their transmitted points therefore \mbox{Algorithm 2} is able to detect them correctly similar to the \ac{ML} estimation.

In Fig. \ref{SER_ML_LD}, the \ac{ML} estimation and \mbox{Algorithm 2} are compared for \ac{QAM} formats in different \ac{SE}s over \ac{AWGN} channel. Interestingly, as the \ac{SNR} (defined as $E_s/(N\cdot N_0 / 2)$) or the number of the constellation points increases, the gap between the \ac{SER} performance of \mbox{Algorithm 2} and the \ac{ML} estimation decreases and they approach each other. Therefore, it can be concluded that, for finite \ac{VC}s, \mbox{Algorithm 2} is close to optimum for constellations with high cardinality, or in the limit of high \ac{SNR}.

\begin{figure*}[h]
\begin{subfigure}{.5\textwidth}
  \centering
  \includegraphics[width=1\linewidth]{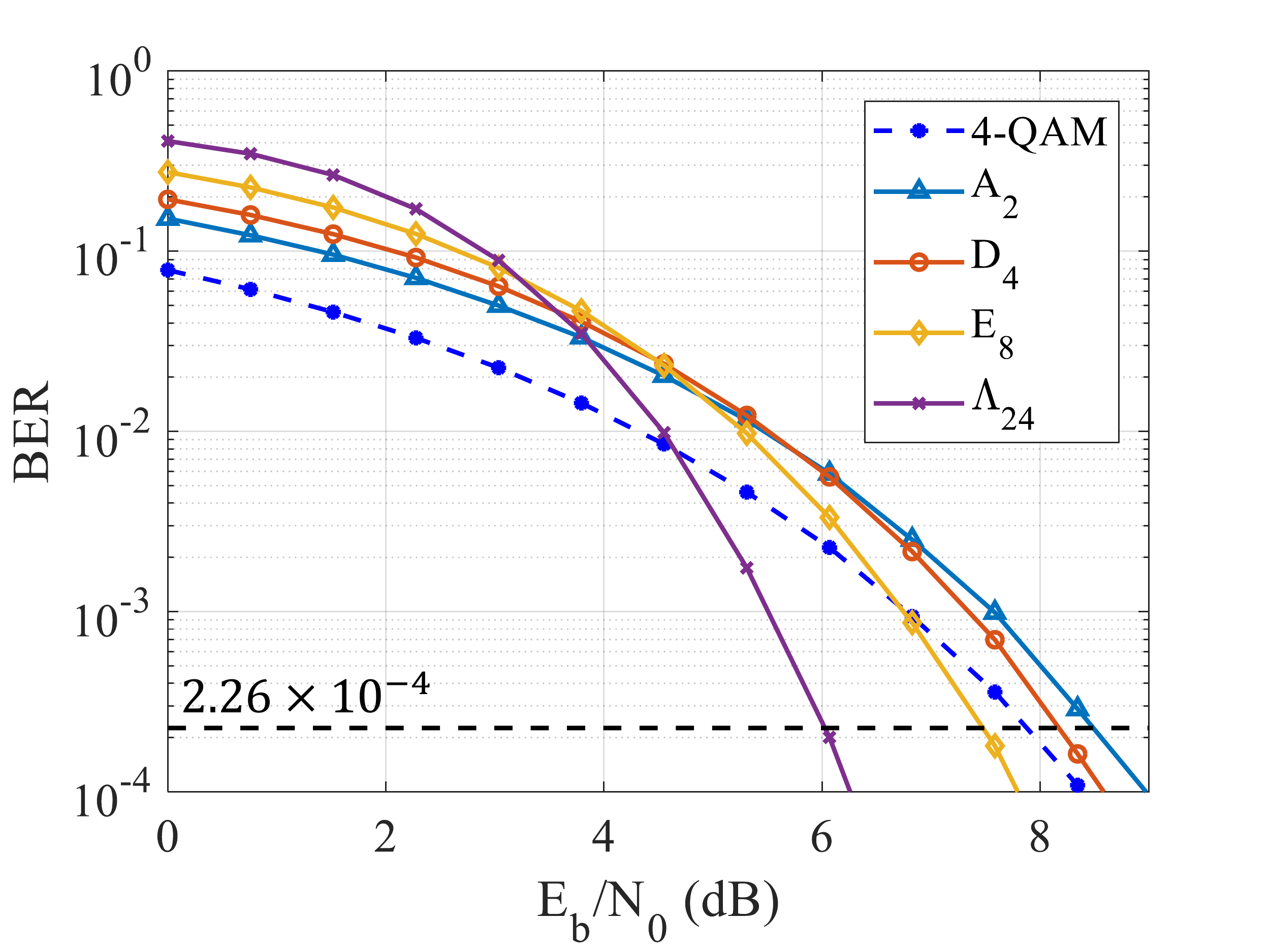}
  \caption{$SE$ = 2 bits/sym/dimension pair}
  \label{SE2}
\end{subfigure}
\begin{subfigure}{.5\textwidth}
  \centering
  \includegraphics[width=1\linewidth]{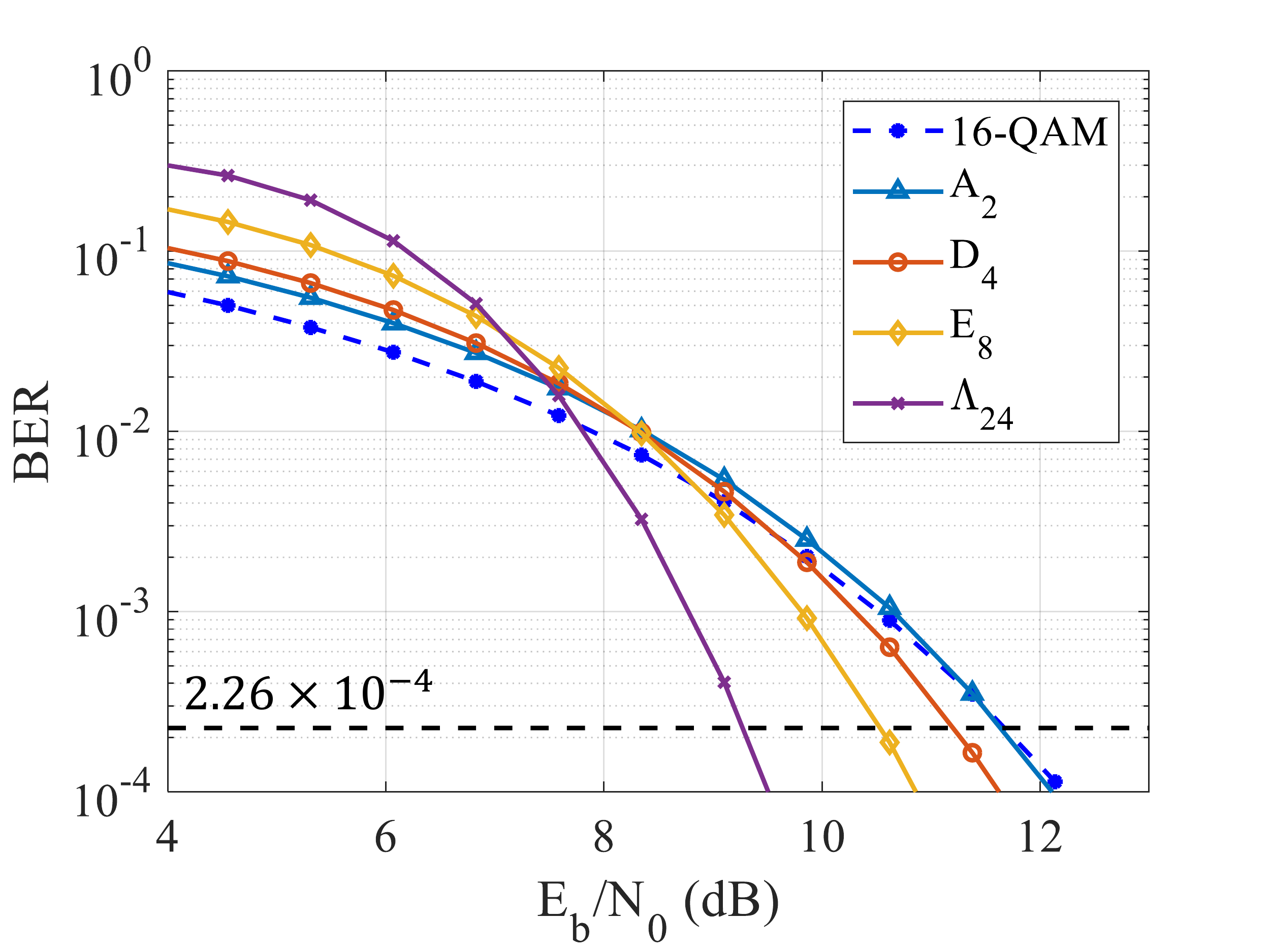}
  \caption{$SE$ = 4 bits/sym/dimension pair}
  \label{SE4}
\end{subfigure}\\
\begin{subfigure}{.5\textwidth}
  \centering
  \includegraphics[width=1\linewidth]{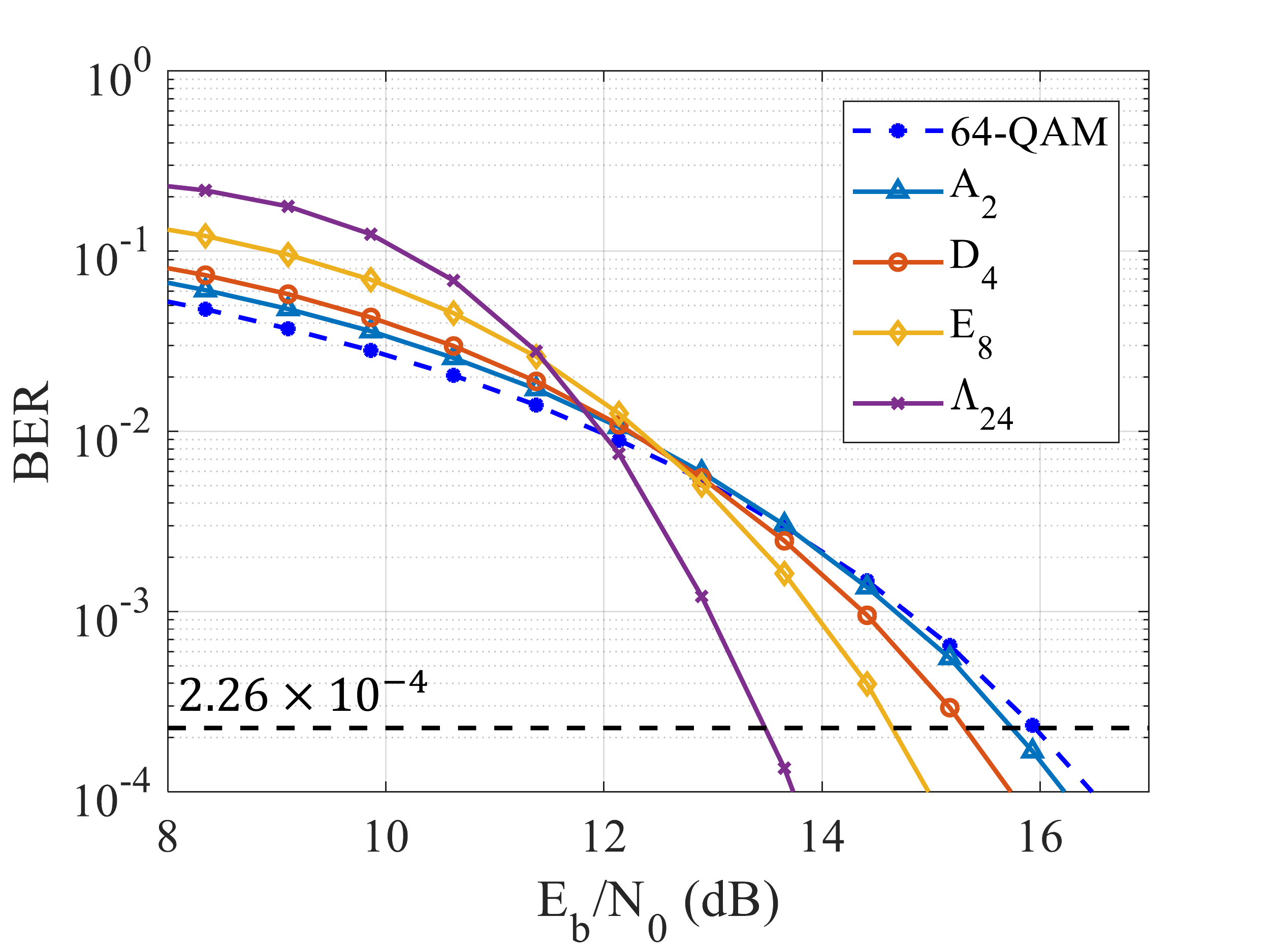}
  \caption{$SE$ = 6 bits/sym/dimension pair}
  \label{SE6}
\end{subfigure}
\begin{subfigure}{.5\textwidth}
  \centering
  \includegraphics[width=1\linewidth]{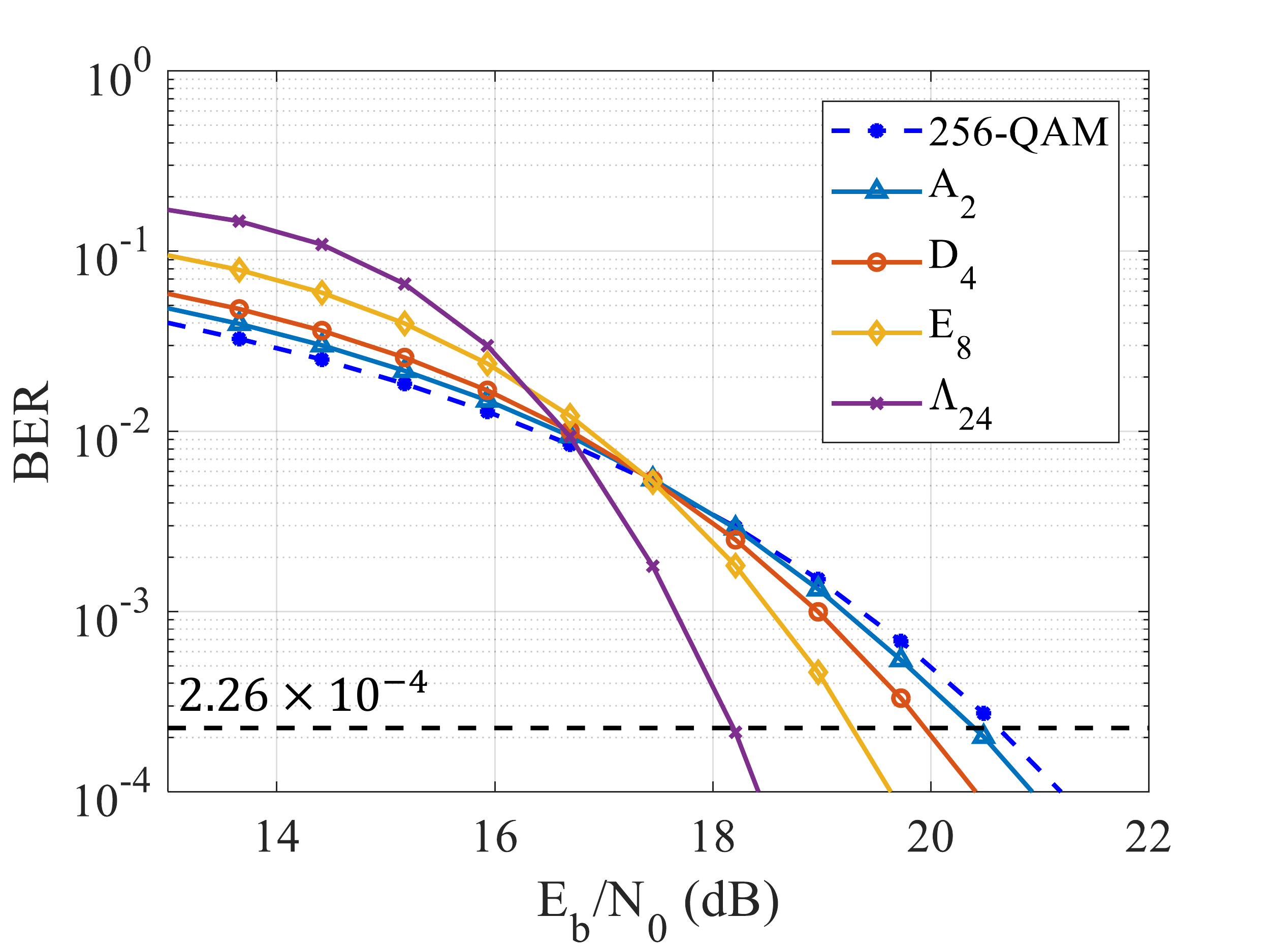} 
  \caption{$SE$ = 8 bits/sym/dimension pair}
  \label{SE8}
\end{subfigure}
\caption{The \ac{BER} vs. $\rm{E_b/N_0}$ for multidimensional modulation formats over \ac{AWGN} channel}
\label{awgn}
\end{figure*}

Finally, in Fig. \ref{awgn}, the \ac{BER} performance of \ac{VC}s in different dimensions are compared with the equivalent \ac{QAM} format at the same \ac{SE}.
Except for the $A_2$ lattice in \mbox{Fig. \ref{SE2}}, all \ac{VC}s will eventually perform better than the \ac{QAM} format as the $E_b/N_0$ increases. The reason for this behavior is the asymptotic power efficiency that was discussed in \mbox{section \ref{figOfMerit}}. However, for low $E_b/N_0$, \ac{QAM} formats outperform \ac{VC}s because of Gray labeling and smaller $\bar{\tau}$.
In \mbox{Fig. \ref{SEvsSP}}, at \mbox{$SE$ = 2 bits/sym/dimension pair}, the $A_2$ and the \ac{QAM} format have the same sensitivity penalty, however, in Fig. \ref{SE2}, 4-\ac{QAM} always performs better than $A_2$ in all $E_b/N_0$. This is because of 4-\ac{QAM} Gray labeling and the average number of closest neighbors which for the $A_2$ lattice with 4 constellation points is $\bar{\tau}=2.5$ and for the 4-\ac{QAM} is $\bar{\tau} = 2$.

\subsection{Nonlinear optical channel}
An important characteristic of the \ac{VC}s is their performance in the nonlinear fiber-optic channel. In order to investigate this problem, a nonlinear fiber-optic channel has been simulated using the split-step Fourier method (SSFM). In an experimental setup, multidimensional modulation formats can be realized in different ways. Among the possible dimensions in fiber, in this work, we focus on in-phase, quadrature, polarization, and wavelength. Using different combinations of dimensions in practice will give different performance results because the coupling and cross-talk between the dimensions in a nonlinear channel might differ.

The simulation setup consists of a transmitter side including a symbol generator that produces multidimensional symbols according to the selected \ac{VC}. These symbols are then parallelized into multiple groups of 4-dimensional symbols because each wavelength can carry up to 4 dimensions in a single-mode fiber. At this stage, each group of 4-dimensional symbols is combined with pilot symbols that have \ac{QPSK} modulation format and they are used for pilot-aided digital signal processing in the receiver. Finally, in the transmitter, the combined payloads and pilots are upsampled and filtered with a \ac{RRC} filter. Then, each group is shifted to its corresponding wavelength and they are added together to form a \ac{WDM} system. Before the channel, the total optical power of the signal is set and then using the Manakov equations \cite{mumtaz2012nonlinear} and SSFM \cite{agrawal2000nonlinear}, the signal is propagated over the nonlinear dispersive fiber-optic channel. The signal is amplified using an erbium-doped fiber amplifier (EDFA) at the end of each span in the fiber link.

On the receiver side, first, digital dispersion compensation is applied for the whole propagation link distance. Then, each wavelength channel is filtered out using an \ac{RRC} filter and downsampled. Using the \ac{QPSK} pilots, polarization demultiplexing and phase tracking are applied to the payload symbols. After that, the processed payloads are combined together to form the multidimensional symbols again and using \mbox{Algorithm 2} and \ac{ML} estimation, it is decided what the transmitted symbol has been for \ac{VC}s and \ac{QAM} format, respectively. In the end, \ac{BER} is calculated based on the transmitted and the received bits. The parameters of the nonlinear simulation setup are listed in Table \ref{nli_sym_param}.
\begin{table}[h]
    \centering
    \caption{Nonlinear simulation setup parameters}
    \begin{tabular}[t]{lccccc}
        \hline
        Parameter&\,&\,&\,&\,&Value\\
        \hline\hline
        Symbol rate&\,&\,&\,&\,&28 GBaud\\
        RRC roll-off factor&\,&\,&\,&\,&0.2\\
        WDM spacing&\,&\,&\,&\,&50 GHz\\
        Fiber nonlinear coefficient &\,&\,&\,&\,&1.3 $\rm W^{-1}km^{-1}$\\
        Fiber dispersion&\,&\,&\,&\,&17 ps/nm/km\\
        Fiber attenuation&\,&\,&\,&\,&0.2 dB/km\\
        Span length&\,&\,&\,&\,&80 km\\
        EDFA noise figure&\,&\,&\,&\,&5 dB\\
        SSFM step size&\,&\,&\,&\,&0.5 km\\
        Oversampling factor&\,&\,&\,&\,&32\\
        Pilot overhead&\,&\,&\,&\,&1.56\%\\
        \hline
    \end{tabular}
    \label{nli_sym_param}
\end{table}%

\begin{figure}
    \begin{center}
        \includegraphics[width=0.5\textwidth]{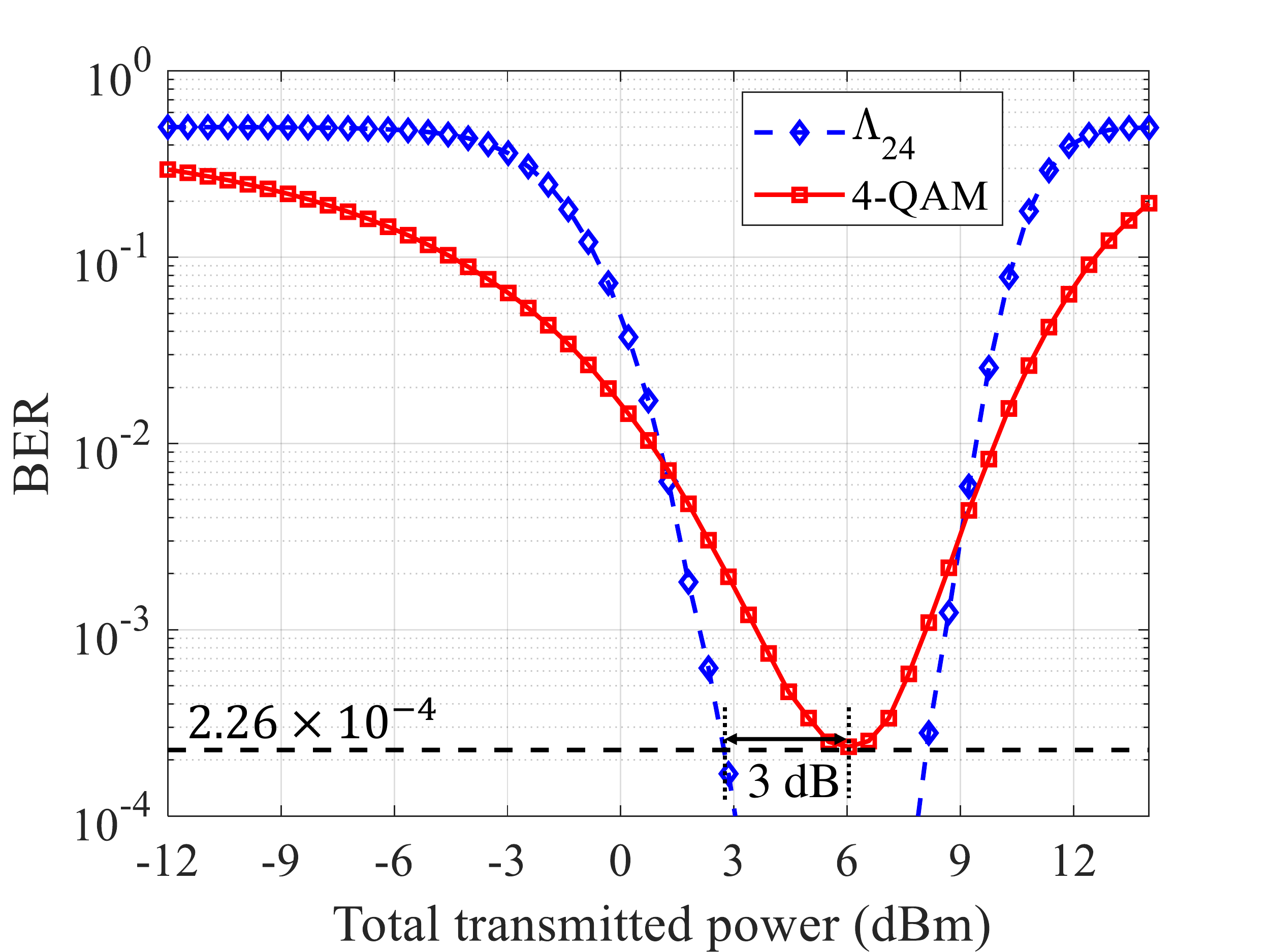}
        \caption{The \ac{BER} vs. total transmitted optical power after 80 spans of 80 km fibers over a WDM system with 6 wavelengths.}
        \label{BER_Pin}
    \end{center}
\end{figure}

In Fig. \ref{BER_Pin}, the \ac{BER} performance of the Leech lattice is compared with the 4-\ac{QAM} at the same \ac{SE}. The total propagation distance is 80 spans (6400 km) using 6 wavelengths to transmit the 24-dimensional signals.
At a \ac{BER} of $2.26\times 10^{-4}$, which corresponds to $10^{-15}$ after error correction using a \mbox{Reed-Solomon (544, 514) (KP4)} code \cite{agrell2018information}, the achieved transmitted power gain is more than 3 dB.

\begin{figure}
    \begin{center}
        \includegraphics[width=0.5\textwidth]{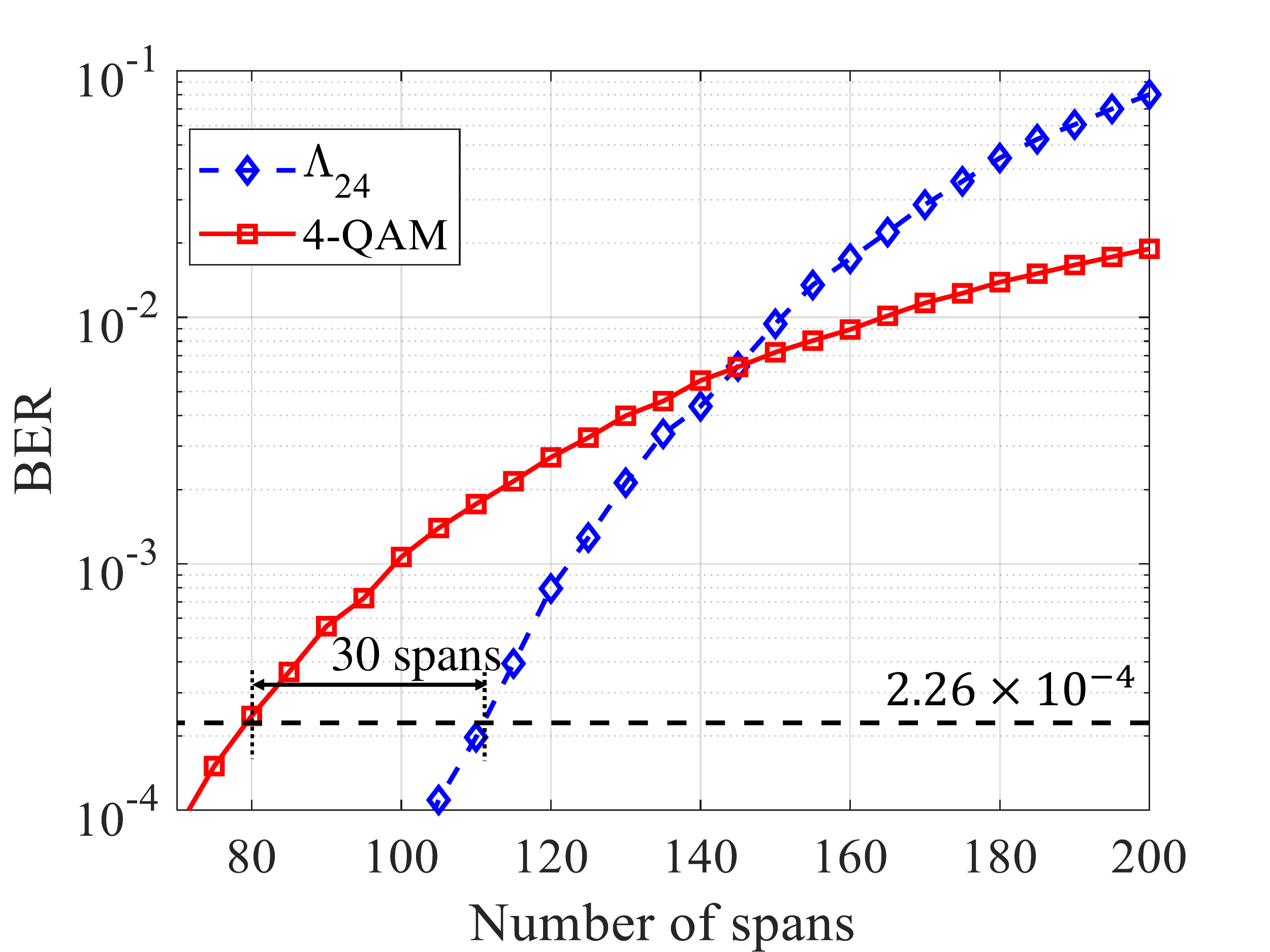}
        \caption{The \ac{BER} vs. transmission distance at the optimum optical power over nonlinear fiber-optic channel.}
        \label{BER_Distance}
    \end{center}
\end{figure}

In \mbox{Fig. \ref{BER_Distance}}, the \ac{BER} performance is shown at the optimum power in different transmission distances. According to our simulations, the optimum power almost remains constant at different transmission distances. In this representation, transmission distance improvement of more than 30 spans or 38\% is shown at a \ac{BER} equal to \mbox{$2.26\times 10^{-4}$}. 

Other shaping methods, e.g., probabilistic amplitude shaping, offer around 7 to 15\% reach improvement depending on the achievable information rate value that they are compared with the uniform \ac{QAM} \cite{fehenberger2016probabilistic,pan2016probabilistic,fehenberger2015ldpc}. Therefore, based on the results of this paper, our geometric shaping method can be considered as a competitive shaping approach with low complexity to increase the performance of optical communication systems.

\section{Conclusion}
We have analyzed Voronoi constellations as a geometric shaping method for optical communication systems. The complexity of modulation and demodulation are low compared with other geometric schemes because modulation and demodulation are carried out algorithmically and the constellation need not be stored in a table, paving the way for more or less unlimited cardinalities. Reach enhancements as high as 38\% is found in nonlinear fiber link simulations.
We conclude that lattice-based geometric shaping can provide higher gains than what has been reported using probabilistic shaping, at much shorter block lengths and moderate complexity.

\section{Acknowledgement}
We wish to acknowledge financial support from the Knut and Alice Wallenberg foundation as well as the Swedish Research Council (VR) in projects \mbox{2017-03702} and \mbox{2019-04078}.
The simulations were performed on resources at Chalmers Centre for Computational Science and Engineering (C3SE) provided by the Swedish National Infrastructure for Computing (SNIC).

\bibliographystyle{IEEEtran}
\bibliography{ref}
\acrodef{QAM}{quadrature amplitude modulate}
\acrodef{VC}{Voronoi constellation}
\acrodef{VCs}{Voronoi constellations}
\acrodef{ML}{maximum likelihood}
\acrodef{CPA}{closest point algorithm}
\acrodef{SNR}{signal to noise ratio}
\acrodef{AWGN}{additive white Gaussian noise}
\acrodef{SER}{symbol error rate}
\acrodef{BER}{bit error rate}
\acrodef{SE}{spectral efficiency}
\acrodef{WDM}{wavelength-division multiplexing}
\acrodef{QPSK}{quadrature phase-shift keying}
\acrodef{BPSK}{binary phase-shift keying}
\acrodef{RRC}{root-raised-cosine}
\end{document}